# Electrochromism of Ni-deficient nickel oxide – Theoretical justification

Igor A. Pašti[1,2]*, Ana S. Dobrota[1], Dmitri Migas,[3,4,5] Börje Johansson,[2,6] Natalia V. Skorodumova[2]*

[1] *University of Belgrade – Faculty of Physical Chemistry, Belgrade, Serbia*

[2] *Department of Materials Science and Engineering, School of Industrial Engineering and Management, KTH – Royal Institute of Technology, Stockholm, Sweden*

[3] *Belarusian State University of Informatics and Radioelectronics, Minsk, Belarus*

[4] *National Research Nuclear University MEPhI (Moscow Engineering Physics Institute), Moscow, Russia*

[5] *Institute for Nuclear Problems of Belarusian State University, Minsk, Belarus*

[6] *Department of Physics and Astronomy, Uppsala University, Uppsala, Sweden*

**\*corresponding authors**

Prof. Igor A. Pašti, e-mail: igor@ffh.bg.ac.rs

Prof. Natalia V.Skorodumova, e-mail: snv123@kth.se

**Abstract**

The development of new electrochromic materials and devices, like smart windows, has an enormous impact on the energy efficiency of modern society. One of the crucial materials in this technology is nickel-oxide. Ni-deficient NiO shows anodic electrochromism whose mechanism is still under debate. Using DFT+$U$ calculations, we show that Ni vacancy generation results in the formation of hole polarons localised at the two oxygens next to the vacancy. Upon Li insertion or injection of an extra electron into Ni-deficient NiO, one hole gets filled, and the hole bipolaron is converted into a hole polaron well-localized at one O atom. Furthermore, the calculated absorption coefficients demonstrate that Li insertion/extraction or rather the addition/removal of an extra electron into Ni-deficient NiO can lead to switching between the oxidized (colored) and the reduced (bleached) states. Hence, our results suggest a new mechanism of Ni-deficient NiO electrochromism not related to the $Ni^{2+}/Ni^{3+}$ transition but based on the formation and annihilation of hole polarons in oxygen p-states.

# 1. Introduction

The ever-increasing energy consumption and associated global environmental problems are urging the rapid development of renewable energy solutions. A better understanding of the mechanisms governing the processes underlying the operation of sustainable energy systems is essential for further progress. One of the issues at hand is the high energy consumption in homes/offices due to heating and cooling. An exciting field in this regard is the electrochromic (EC) materials, whose optical properties can be controlled reversibly between the transparent (bleached) state and the colored state by applied DC voltage. These materials include some metal oxides, conductive polymers, and some inorganic non-oxides mixed ionic and electronic conductors [1]. EC materials can be integrated into multilayer electrochromic devices (ECDs) [2]. Among EDCs, smart windows are of special interest for winning the energy battle since their development may significantly reduce heating/cooling energy consumption [3], especially when combined with photovoltaics [4].

Electrochromism is generally associated with the insertion/extraction (intercalation/de-intercalation) of small ions (such as $H^+$ or $Li^+$) into and out of the EC material layers [3,5]. EC materials are divided into cathodic and anodic ones, defined as those that get colored under ion insertion and the ones coloring under ion extraction, respectively [6]. The most commonly used cathodic EC oxides are mainly based on tungsten oxides, niobium oxide and titanium dioxide, while the anodic ones are mostly based on nickel oxide [1]. In particular, NiO turns from transparent to dark brown upon applying a voltage of about 1.35 V *vs.* Reversible Hydrogen Electrode (RHE) [3]. NiO is especially convenient since it is cheap, abundant, easily fabricated [3], and exhibits relatively good cyclic reversibility [1]. The electrochromism of NiO films prepared by different methods (including CVD, sputtering, spray pyrolysis, sol-gel and electrodeposition) has been demonstrated by many authors [3,7–10]. In recent years different strategies aiming to improve the EC performance of NiO have been proposed, such as the formation of Li-doped NiO films by pulsed layer deposition [2,11,12], formation of double-layer amorphous $NiO_x$ films [10], or post-synthetic ozone treatment of Al-containing NiO [11]. Nonetheless, the mechanism of electrochromism in NiO is still to some extent unresolved, especially in Li-based electrolytes [13]. Since EC materials are mixed conductors, ions can be quickly and reversibly moved, in step with the electrons. These materials experience both chemical and optical changes by simultaneous ion and electron injection/extraction [14]. In Goodenough's model, the insertion of electrons has the sole role of shifting the Fermi energy [5]. However, it is generally accepted that in alkaline media the coloration results from the $Ni^{2+}$ to $Ni^{3+}$ redox change [3] upon ion extraction. Granqvist has noticed that ion transport is not crucial for the basics of electrochromism and that the electrons that enter the material in conjunction with the ions (to obtain overall charge neutrality), in fact, cause the optical effect [5].

Electrochromism has been documented in hydrated nickel oxide, with the proposed EC reaction being: $[Ni(OH)_2]_{bleached} \leftrightarrow [NiOOH + H^+ + e^-]_{colored}$, where coloring is associated with hydrogen extraction [15]. In Li-containing electrolytes, anodic electrochromism of NiO was ascribed to the $Ni^{2+}$-$Ni^{3+}$ transitions during $Li^+$ insertion. According to Passerini *et al.* [16–18], the as-deposited NiO film is initially bleached following the reaction $NiO_x + yLi^+ + ye^- \rightarrow Li_yNiO_x$. The lithiated phase then undergoes reversible transitions between the colored and bleached state as $Li_yNiO_x$ (bleached) $\leftrightarrow Li_{(y-z)}NiO_x$ (colored) $+ zLi^+ + ze^-$. Another proposed mechanism involves transition between transparent $Li^+_{2x}Ni^{2+}_{(1-x)}O^{2-}$ and colored $Li^+_{(2x-y)}Ni^{2+}_{(1-x-y)}Ni^{3+}_yO^{2-}$, as $Li^+_{2x}Ni^{2+}_{(1-x)}O^{2-} \leftrightarrow Li^+_{(2x-y)}Ni^{2+}_{(1-x-y)}Ni^{3+}_yO^{2-} + yLi^+ + ye^-$ [19]. Wen *et al.* have studied the well-characterized sputter-deposited NiO films ($NiO_x$, $1.16 \leq x \leq 1.32$) in $LiClO_4$ in propylene carbonate electrolyte [13]. They found that the optical modulation range was enhanced with increasing $x$ and as the voltammetric cycling potential window was widened [13]. Their results are consistent with the model assuming both cations and anions contributing to the charge exchange and electrochromism through surface rather than bulk reactions [13]. This is particularly important as the different proposed mechanisms of NiO electrochromism involved both surface [20] and bulk species formation [19]. In fact, it has been concluded that also the surface orientation has a significant effect on electrochromism of NiO, suggesting that high density of Ni sites on (111) and small interatomic distances, compared to those of the (001) surface, give enhanced optical absorption due to $H^+$ and $OH^-$ adsorption/desorption [7].

Although discovered over thirty years ago, the underlying mechanism of NiO coloration/bleaching is still under active debate. Deeper, atomic-level comprehension of this mechanism could provide a pathway for further improvement of EC properties of nickel oxide. It could also offer general guidelines for the design of other promising EC materials. Here we employ the Density Functional Theory with on-site Coulombic interactions (DFT+*U*) calculations to investigate the electrochromism in NiO. This work focuses on bulk NiO, allowing us to exclude any effects of surface adsorption processes involving ions as well as surface orientation. Nevertheless, we observe a clear modulation of optical spectra, which offers an alternative view on the electrochromism of NiO.

## 2. Methods

The first-principle DFT calculations were performed using the Vienna ab initio simulation code (VASP) [21–24]. The Generalized Gradient Approximation (GGA) in the parametrization by Perdew, Burk and Ernzerhof [25] combined with the projector augmented wave (PAW) method was used [26]. The cut-off energy of 800 eV and Gaussian smearing with the width of $\sigma$ = 0.025 eV for the occupation of the electronic levels were used. The on-site Coulomb interactions were added to the *d*-states of Ni as well as to the *p*-states of O using the DFT+*U* scheme of Liechtenstein *et al*. [27]. Non-spherical contributions from the gradient corrections inside the PAW spheres and Kohn-Sham potential were accounted for. Monkhorst-Pack *k*-point mesh was employed for the relaxation calculations [28]. For the Density of States (DOS) calculations, the Blöch tetrahedron method was used [29]. The calculations were done using two diffretent NiO supercells: rhombohedral cell built up as a 1×2×2 multiply of primitive cell, containing 16 atoms, and a larger hexagonal supercell with alternating Ni/O layers containing 108 atoms. For each of them, the convergence with respect to the *k*-point mesh was carefully checked. The relaxation proceeded until the Hellmann-Feynman forces acting on all the atoms became smaller than $5\times10^{-3}$ eV Å$^{-1}$. Spin-polarization was taken into account in all the calculations. Optical spectra were calculated from the frequency-dependent microscopic polarizability matrix in the projector-augmented wave (PAW) methodology [30]. VESTA code was used for visualization and the simulation of X-ray diffraction patterns [31]. Partial atomic charges were evaluated using Bader analysis [32]. The magnetic structure of pristine NiO bulk was antiferromagnetic, where the alternating (111) layers of Ni$^{2+}$ ions bear opposite spin orientations. For data analysis and graphical presentation VASPKIT [33] and VESTA code [34] were used.

## 3. Results and Discussion

### 3.1 Hybrid *vs*. DFT+*U*: unit cell calculations

NiO is a typical example of a correlated material where conventional DFT approximations fail to explain the electronic structure properly. For this reason, modelling of NiO requires approaches going beyond standard DFT, at least at the level of hybrid potentials or DFT+*U*. However, hybrid calculations are extremely time-consuming, and for large cells they are truly heavy to perform. DFT+*U* is much more time-efficient while the addition of the on-site Coulomb interaction improves the description of the electronic structure being often able to reproduce the essential features in good agreement with hybrid calculations. Hence, here we directly compare the results obtained by DFT+$U$ and HSE for the unit cell of NiO and in the following for the larger supercell we use DFT+$U$ only.

The NiO lattice optimization was done using the 1×1 hexagonal cell with alternating Ni and O layers, taking into account the G-type of antiferromagnetic ordering in NiO. For different values of $U$ applied to the Ni $d$- and O $p$-states we performed structural optimization to determine the equlibrium lattice constant, magnetic moment and the bandgap. As a reference, we also performed hybrid HSE06 calculations with the 800 eV cut-off energy, in which we obtained the lattice constant of 1.0019×$a_{0,exp}$ ($a_{0,exp}$ = 4.17 nm), the magnetic moment on Ni atom of 1.653 $\mu_B$ and the bandgap of 4.2 eV. The experimentally reported bandgap of NiO is in between 3.4 eV and 4.3 eV, depending on the actual method used to determine it [35]. The value of the bandgap is also found to depend on the stoichiometry of NiO [36]. In particular, Becker *et al*. [36] have reported values between 3.35 and 3.6 eV with the maximum near the 1:1 stoichiometry.

We have tested different values of $U$ applied to the $d$-states of Ni and found that $U$ between 4 and 8 eV ($J$ = 1 eV) increases the bandgap from less than 1 eV (DFT calculations) to the values between 2.9 (for $U$ = 4 eV) and 3.6 eV (for $U$ = 8 eV). Simultaneously, the magnetic moment on Ni atoms increases with the addition of $U$ from 1.3 $\mu_B$ (DFT calculations) to 1.65 (for $U$ = 4 eV) and 1.80 $\mu_B$ (for $U$ = 8 eV) (see Supplementary Information, **Fig. S1**). The lattice parameter increases slightly, from 1.013 to 1.017 $a_{0,exp}$, when $U$ at Ni $d$-states increases from 4 to 8 eV. Comparing the experimental parameters and the calculated ones, we have chosen $U$ = 6.5 eV for Ni with $J$ = 1 eV. Using this set of $U$ and $J$, we obtained the best agreement between the lattice parameter, magnetic moment of Ni and bandgap (Supplementary Information, **Figs. S1-S3**) with the experimental data and the HSE06 calculations. The overall features of the NiO density of states obtained using DFT+$U$ and hybrid calculations are in good agreement. Moreover, $U$ = 6.5 eV agrees well with the values of $U$ previously reported in the literature [37].

Additionally, we considered the effect of the $U$ addition to the $p$-states of O. The application of $U$ to the $p$-states of O causes further decrease of lattice constant and increase of the bandgap, as shown in **Table 1**. It also leads to some increase of partial charges on O atoms indicating better charge localization. Finally, we chose the value of $U$ = 9 eV ($J$ = 1 eV) for O p-states, in agreement with previous works on charge localization in oxide materials [38]. DOSs calculated with the chosen set of $U$ parameters are presented in **Fig. 1** and compared to the results of our hybrid calculations.

**Table 1.** The lattice parameters ($a/a_{0,\text{exp}}$), magnetic moments on Ni atoms ($\mu$(Ni)), charge on O atoms ($Q$(O)) and the values of the bandgap of NiO, depending on the value of $U$ added on the $p$-states of O, with $U$ on Ni $d$-states being 6.5 eV and $J$ = 1 eV in all cases.

| U@Ni / eV | U@O / eV | $a/a_{0,\text{exp}}$ | $\mu$(Ni) ($\mu_B$) | Q(O) (e) | Band gap (eV) |
|---|---|---|---|---|---|
| 6.5 | 0 | 1.0156 | 1.741 | 7.245 | 2.997 |
| 6.5 | 5 | 1.0128 | 1.748 | 7.255 | 3.203 |
| 6.5 | 9 | 1.0096 | 1.752 | 7.268 | 3.463 |
| 6.5 | 13 | 1.0062 | 1.757 | 7.282 | 3.703 |

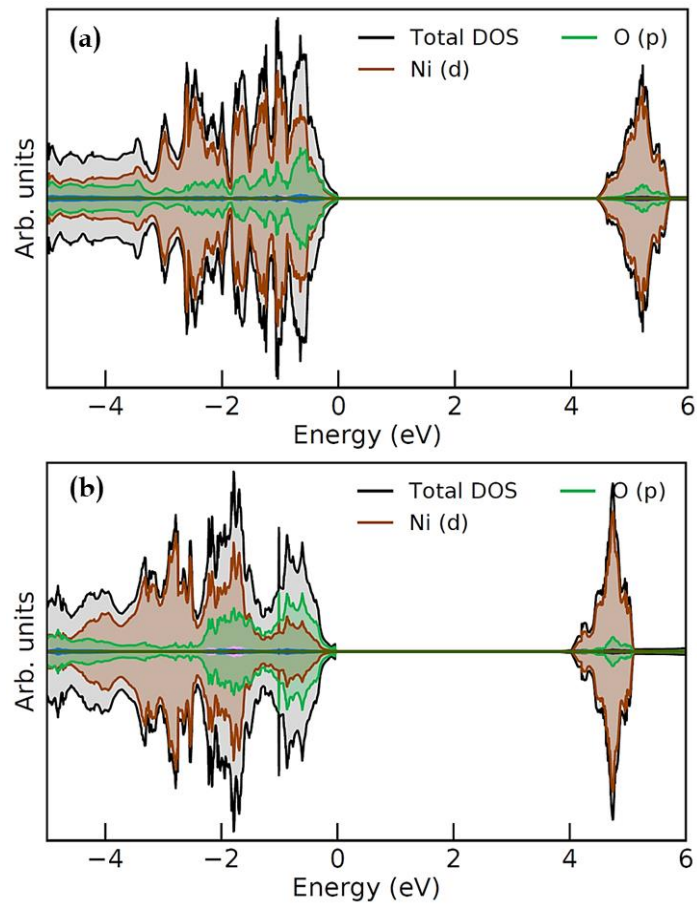

**Figure 1.** DOS of pristine NiO calculated using HSE06 hybrid calculations (a) and DFT+$U$ calculations (b) with $U$ on Ni $d$-states of 6.5 eV and $U$ on O $p$-states of 9 eV ($J$ = 1 eV in both cases).

3.2 Ni and O vancacy formation (16 atom cell)

As the next step we consider the formation of Ni and O vacancies in NiO, which we first do using a rhombohedral cell built up as a 1×2×2 multiply of the NiO cubic unit cell, which contained 16 atoms in total. This cell was chosen because the formation of one O or Ni vacancy in it results in the formal stoichiometry of NiO$_{0.85}$ or NiO$_{1.14}$. The latter is particularly interesting as it is close to the NiO stoichiometry experimentally investigated by Wen *et al*. [13]. It is also important to note that highly Ni-deficient NiO retains the NiO structure [13], as shown by X-ray diffraction, although some variations in relative intensities of different reflections have been observed. The calculated DOS and simulated X-ray diffraction patterns of NiO$_{0.85}$ and NiO$_{1.14}$ are presented in **Fig. 2**.

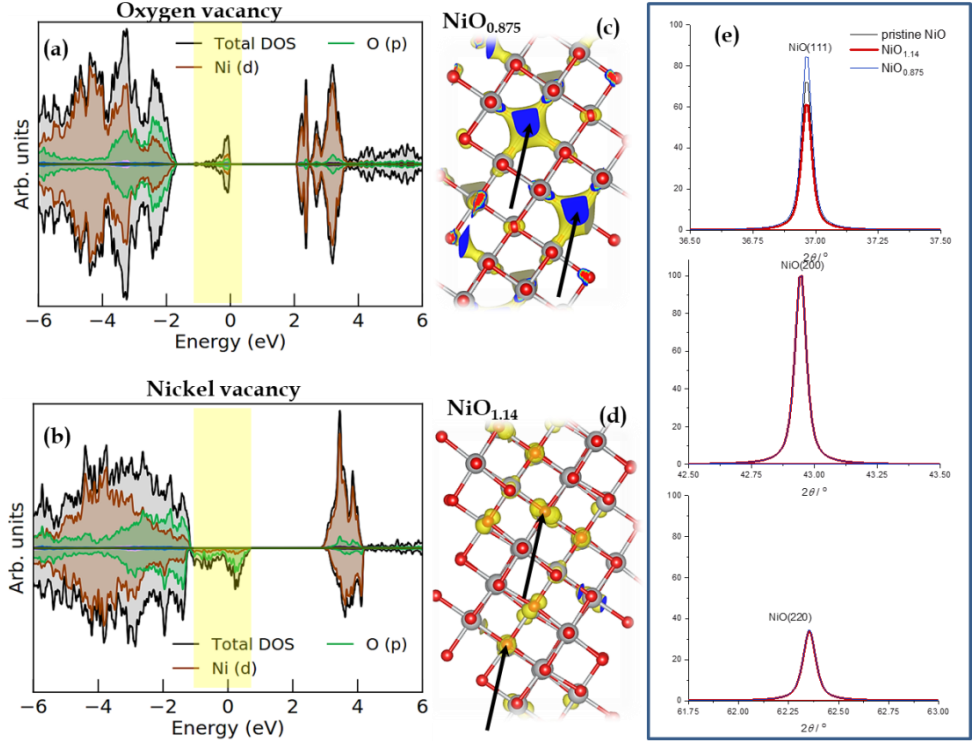

**Figure 2.** DOS plots for O-deficient (a) and Ni-deficient NiO (b). In the middle, integrated partial local densities of states in the regions marked in DOS plots are presented for O-deficient (c) and Ni-deficient NiO (d). The integration was done in the marked energy window. On the right, the simulated XRD patterns of O-deficient and Ni-deficient NiO are compared with the X-ray diffraction pattern of pristine NiO (e; (111), (200), and (220) reflections are shown). Both O- and Ni-deficient systems show no magnetization with the local moments being AFM arranged. Arrows indicate oxygen (NiO$_{0.875}$) and nickel (NiO$_{1.14}$) vacancies sites.

Even in the small simulation cell (1×2×2, 16 atoms), DOS and the integrated densities of states (ILDOS) indicate charge localization in both O- and Ni-deficient NiO. In the case of oxygen vacancy we observe the formation of electronic polarons at two neighbouring Ni atoms arising from the localization of two electrons left after the removal of a neutral O atom (**Fig. 2**, a and c). Here, however, we are more interested in Ni-deficient NiO, as Ni-deficient NiO is electrochromic [4], so we study it in more detail. The Bader charges obtained for the O atoms in NiO$_{1.14}$ show that all the Ni atoms have the same charge (+1.28e). However, O atoms do not have uniform charges as a hole bipolaron forms in the NiO$_{1.14}$. The polaron states are located just above the Fermi level and they are merged with the valence band. We notice that in the case of the small supercell we do not see full localization at two oxygen atoms due to size constrain. In the following we will called this situation "partially delocalized". The empty states are clearly visible in **Fig. 2** (b) and they are predominantly oxygen states. This conclusion is very important as we see that all the Ni atoms are basically in the same oxidation state in Ni-deficient NiO. Our finding agrees well with the computational results of Ferrari *et al*. [39] who used both DFT+*U* and Becke hybrid-exchange functional with 20% of exact Fock exchange, together with the Perdew-Wang correlation functional (B3PW). Their hybrid calculations showed that the holes were localized at the oxygen ions nearest to the vacancy. However, their DFT+*U* calculations resulted in substantional delocalization, thus failing to localize holes at two O atoms [39].

3.2. Charge localization in Ni-deficient NiO (108 atom cell)

As we have demonstrated in the previous section there is a clear indication of the occurrence of charge localization in Ni-deficient NiO. The use of a relatively small simulation cell makes it difficult to observe these effects clearly. For this reason, we performed another set of calculations using a larger hexagonal cell of 108 atoms. A simple removal of one Ni atom from such a simulation cell (giving the stoichiometry of $NiO_{1.019}$) and straightforward relaxation provides a partially delocalied hole distribution (**Fig. 3**) similar to that obtained in the smaller cell (**Fig. 2**). Again, all Ni atoms have the same partial charge (oxidation state).

It is well-known, however, that an accurate account of local distortions around the site where polaron is forming is crutial for the correct description of the phenomenon. Having taken special care about local distortions in the vicinity of the Ni vacancy we obtained a more stable, solution, showing a clear formation of the hole bipolaron at two neighbouring O atoms (**Fig.3**). This solution is by 0.41 eV more stable than the partially delocalized one. We note that the system with the bipolaron shows no magetisation as the removal of one spin-down Ni is compensated by the spin-down orientation of the polaron magnetic moments. In fact, we managed to stabilise some other magnetic configutions as well but the non-magnetic one proved to have the lowest energy.

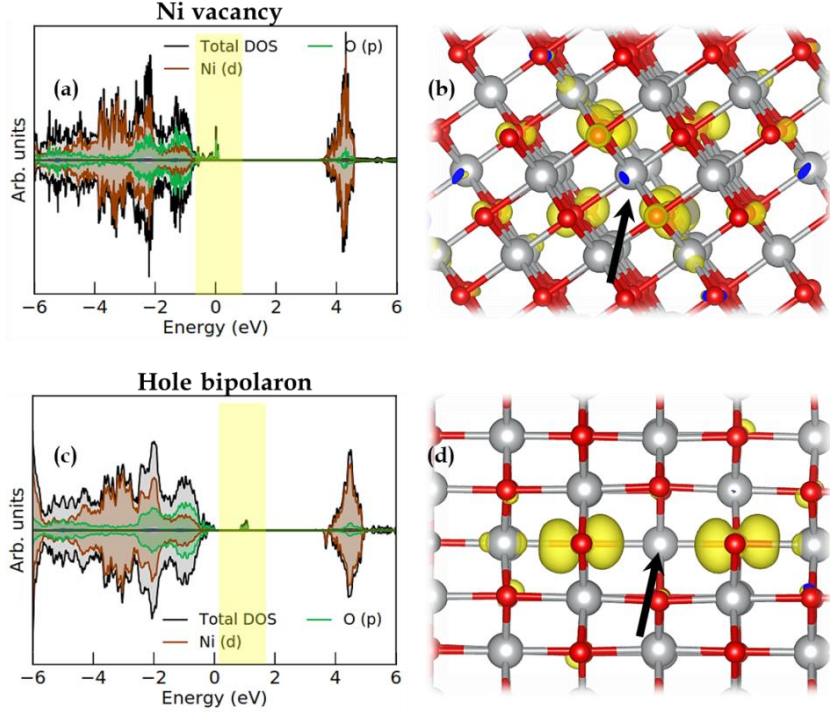

**Figure 3.** DOS and ILDOS of Ni-vacancy in NiO$_{1.019}$: upper row presents a partially delocalized solution (DOS, a, and integrated partial density of states, b) while the bottom row presents a hole bipolaron located at O *p*-states (DOS, c, and integrated partial density of states, d). The integration was done in the marked energy window. Arrows indicate oxygen and nickel vacancy sites.

3.3. Electrochromism of NiO

As the majority of experimental data is about the electrochromism of NiO in Li-containing electrolytes, we investigated the insertion of a Li atom (that is Li$^+$ + e$^-$; in an experiment, the electron comes from the external circuit) into the Ni vacancy. We checked that the insertion of Li into the Ni vacancy in NiO$_{1.14}$ (small cell) is an exothermic process ($\Delta E$ = −5.5 eV, for the reaction, more below), while the Li insertion into the O vacancy is endothermic ($\Delta E$ = +1.1 eV), suggesting that Ni vacancies can be probable sites for Li incorporation in NiO. As we work with bulk here, we note that Li insertion and diffusion in NiO bulk was recently investigated by Luo *et al*. [40], showing that the Li$^+$ diffusion in NiO bulk was feasible even for pristine NiO. However, we do not expect qualitatively different behavior whether Li$^+$ + e$^-$ interacts with the bulk or surface Ni vacancy. Moreover, we note that Li+ intercalation into Ni-deficient NiO occurs at potentials above +3 V *vs*. Li$^+$/Li in non-aqueous Li-containing electrolytes [13], which is in line with the energy of Li insertion into Ni vacancy, taking into account that cohesive energy of Li is 1.63 eV atom$^{-1}$ [41].

First, we place one Li atom into the Ni vacancy formed in the small cell (16 atoms cell, $NiO_{1.14}$ stoichiometry) following the reaction $0.143Li^+ + NiO_{1.14} + 0.143e^- \leftrightarrow Li_{0.143}NiO_{1.14}$. This process corresponds to the reduction of $NiO_{1.14}$ to $Li_{0.143}NiO_{1.14}$. The result of this process is the filling of one hole with an electron coming from the Li atom so that only one well-localized hole polaron remains in the system. The polaron states are localized in the bandgap being well-separated from the valence band (**Fig. 4**). We calculated the optical spectra for the cells with and without Li to see if there is any difference. The comparison of the optical spectra calculated for oxidized $NiO_{1.14}$ and reduced $Li_{0.143}NiO_{1.14}$ show a decreased absorption coefficient in the visible region upon Li insertion (**Fig. 4**), i.e., upon the reduction of Ni-deficient NiO.

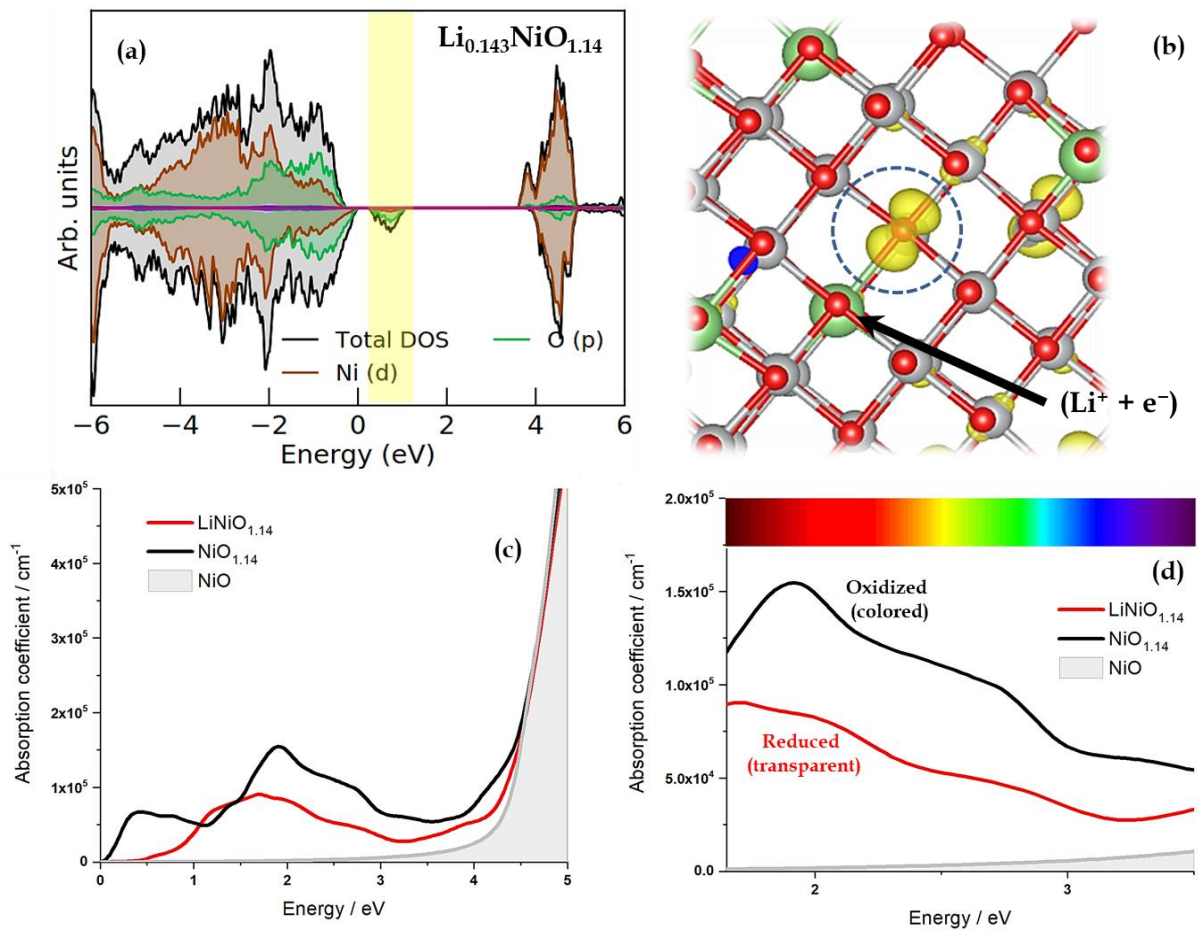

**Figure 4.** Density of States (a) and integrated local density of states (b) of $Li_{0.143}NiO_{1.14}$. The bottom row shows electronic spectra of pristine NiO, $NiO_{1.14}$ and $Li_{0.143}NiO_{1.14}$ in a wide energy range (c) and the visible region (d).

The absorption spectra presented in **Fig. 4** show that pristine NiO is practically completely transparent up to the photon energies of approx. 4 eV, where a steep increase of the absorption coefficient is seen due to the interband transitions. For $NiO_{1.14}$, we observe free carrier absorption since some valence bands are above the Fermi level due to the partial delocalization of the hole states (**Fig. 2**, up to 1 eV above the Fermi level).

Following the calculated optical spectra of $NiO_{1.14}$ and $Li_{0.143}NiO_{1.14}$, it can be concluded that the reduced state corresponds to transparent state while the oxidized state corresponds to the colored state. We note that the calculated absorption coefficients are in the order of $1 \times 10^5$ cm$^{-1}$ agreeing with experimental results reported in Ref. [42]. It is important to note that in the bleached state, the calculated absorption coefficient does not drop down to zero, in agreement with the results of Wen *et al*. [13], who showed that for the $NiO_x$ films (1.16 < $x$ < 1.32) transmittance changed between, on average, ~85% for the bleached state and ~55% for the colored state.

We also analyzed the effect of Li intercalation into the Ni vacancy for $NiO_{1.019}$ (larger hexagonal cell) and we came to the same conclusion as in the case of $NiO_{1.14}$. Since in the large cell hole bipolaron state is well-separated from the NiO valence band being located in the bandgap (**Fig. 3**), the insertion of ($Li^+$ + $e^-$) results in filling one hole leaving only one hole polaron in the system located at the oxygen next to the Li ion (**Fig. 5**). As in the case of the smaller cell, Li insertion causes the absorption coefficient to decrease, so the resulting oxidized state corresponds to the colored form of NiO. The calculated absorption coefficients for $NiO_{1.019}$ are lower than those for $NiO_{1.14}$ (**Fig. 4**), indicating that the hole localization in the O 2*p* valence band is responsible for the optical spectra changes. As the hexagonal symmetry of NiO is broken by hole bipolaron formation and Li insertion in the Ni vacancy, the diagonal elements of absorption coefficient are different (Supplementary information, **Fig. S4**). However, due to the localization of the polaron states in the bandgap, the concentration of free carriers is low thus the adsorption in the infrared region is low, in contrast to the absorption of of $NiO_{1.14}$ (**Fig. 5, c**).

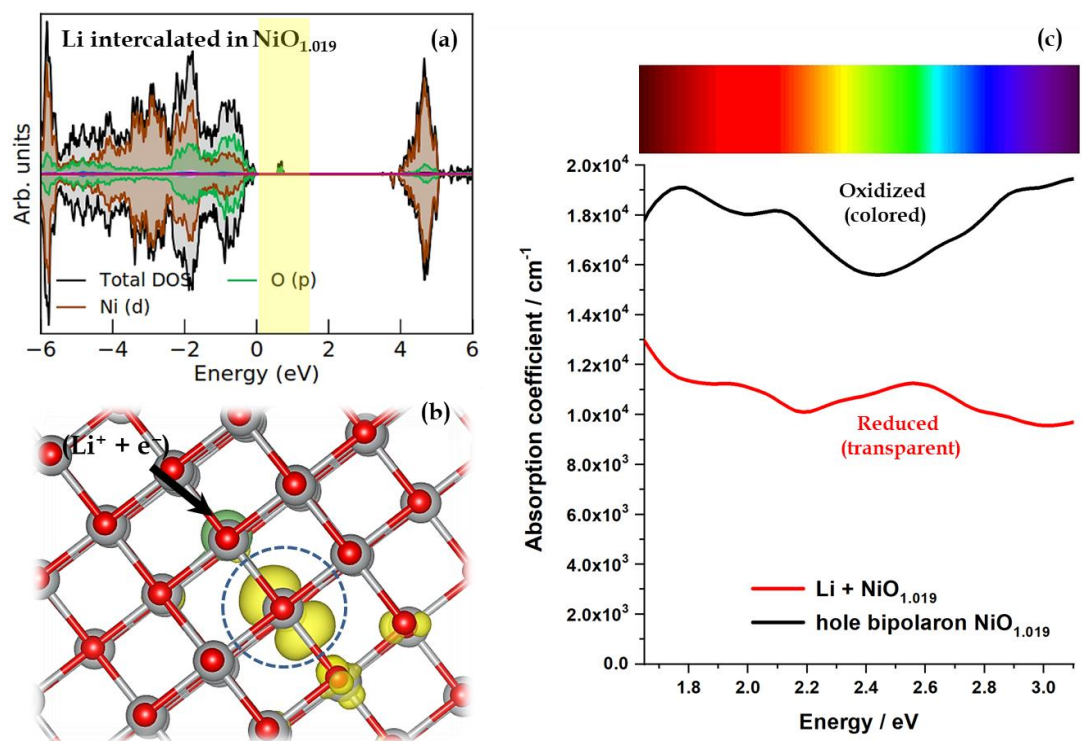

**Figure 5.** DOS (a) and ILDOS (b) of Li intercalated in $NiO_{1.019}$. On the right, optical spectra of $NiO_{1.019}$ before and after Li intercalation are compared (c). Compared to the 16-atoms cell case ($NiO_{1.14}$), the localized hole polaron state is approximately the same energies but narrower.

Filling of the hole bipolaron states was also confirmed by integrating these states, showing that these states could occupy two holes before Li insertion, while after Li insertion, only one hole is left. We note that the charge transfer from Li to O atoms in NiO was also observed by Luo *et al*. [40] using the charge difference analysis in the case of Li adsorption on the O-terminated NiO(111) surface. However, the authors did ascribe the optical modulation of NiO to the transition between Ni(II) and Ni(III) states as the XPS analysis showed the change of the $Ni^{2+}/Ni^{3+}$ ratio in the bleached and colored state. Our results, however, suggest that the change of the optical properties of Ni-deficient NiO should not essentially be linked to the Ni(II)-Ni(III) transition but it can instead be related to the filling of the polaronic states appearing due to the Ni vacancy formation. Further information about the partial charges can be found in **Section S3**, Supplementary information.

NiO electrochromism was observed in different electrolytes. In all of them, however, the ion intercalation is accompanied by the electron injection into the material. To confirm the idea that not ions but the electrons play the crucial role in the electrochromic phenomena we performed calculations for charged Ni-deficient NiO cells (i.e., cells with one extra electron added). We carried out such calculations for the small 16 atom cell and obtained very similar results to those we had for the Li intercalation case. The comparison of DOSs calculated for the "charged" system (**Fig. 6**) and $Li_{0.143}NiO_{1.14}$ (**Fig. 4,** a) demonstrates that the hole polaron states are present above the Fermi level in both cases. One hole is found upon the integration of these states for the charged cell as well as it was for the Li intercalation case. And again, the hole polaron formation lowers the energy of the system by 0.12 eV as compared to the case of partial delocalisation.

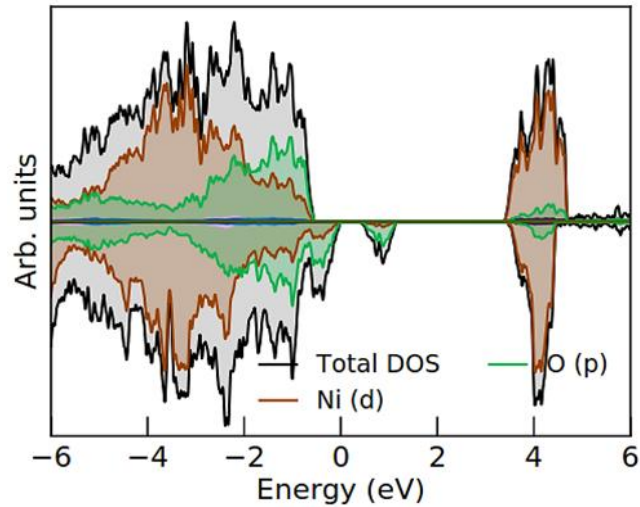

**Figure 6.** DOS of charged $NiO_{1.14}$ when the hole polaron is formed upon adding one electron to the system. This result should be compared with the Li insertion in $NiO_{1.14}$ (**Fig. 4,** a) where almost identical DOS is obtained.

During the operation of an electrochromic device, electrons come from the outer electrical circuit, while ions present in the electrolyte compensate for this charge and take care of electroneutrality. As the Ni-deficient NiO charging has the same effect on the electronic structure (which ultimately shows in the optical properties) as Li insertion, we believe that these results explain why Ni electrochromism is observed in different electrolytes and not only in Li-containing ones. In other words, the charge transfer process itself is responsible for the color change. Different counter ions simply serve to provide charge neutrality of NiO. However, due to different sizes, these counter ions can cause different lattice distortions, affecting optical properties to a certain extent, or lead to irreversible structural changes, or simply cannot accommodate at the vacancy site, preventing or significantly weakening electrochromic effect. This scenario is in line with current opinion that the reversible charge-transfer process between $Ni^{2+}$ and $Ni^{3+}$ is responsible for electrochromism, that is, the extraction and insertion of 3d electrons of Ni in valence band [42–44]. However, while our results also suggest that electron exchange can be responsible for electrochromism, the focus is shifted to the filling of polaronic states which can be responsible for the modulation of optical properties. Thus, we offer a new physical mechanism of the NiO optical property modulation. We notice, however, that the chemical changes occurring in different electrolytes can trigger different degradation mechanisms, passivation, and so on, ultimately influencing the performance and stability of electrochromic NiO films.

Although the presented results give an alternative view on the NiO electrochromism mechanism to that associated with the change of the $Ni^{2+}/Ni^{3+}$ oxidation state, which according to some authors, happens predominantly on the surface [13], these mechanisms do not necessarily exclude each other. The oxidation process during potential cycling is taking place predominantly in the surface layer and possibly a few subsurface layers. This process is yet to be addressed in detail, but the overall results clearly suggest that the physics of NiO electrochromism is much more complex than initially considered.

## 4. Conclusions

We have analyzed the origin of the electrochromism of Ni-deficient NiO using DFT+$U$ calculations. The presented findings suggest an alternative mechanism of NiO electrochromism to the traditional opinion that the change of the optical spectra occurs due to the change of the $Ni^{2+}/Ni^{3+}$ oxidation state. Our results demonstrate that a hole bipolaron formation takes place in Ni-deficient NiO. When Li is inserted into the Ni vacancy, one hole state gets filled, and only one hole polaron is left in the system. The electronic structure changes directly show in the optical spectra, making the absorption coefficient of reduced (Li-intercalated) NiO lower than that of initial Ni-deficient NiO. The obtained results suggest that during the Li insertion process (reduction), the oxidation states of Ni atoms do not change noticeably. We suggest that this process is quite general, and the hole bipolaron states get filled by the simple addition of one electron to the system, being in this sense indeferent to specific positive counter ions. This might explain why Ni electrochromism has been observed in different electrolytes, and not only in those Li-containing.

**Acknowledgment**
I.A.P and A.S.D acknowledge the support provided by the Serbian Ministry of Education, Science and Technological Development (451-03-9/2021-14/200146). The computations and data handling were enabled by resources provided by the Swedish National Infrastructure for Computing (SNIC) at the NSC center of Linköping University, partially funded by the Swedish Research Council through grant agreement no. 2018-05973.

# SUPPLEMENTARY INFORMATION

## S1. Setting up computational parameters

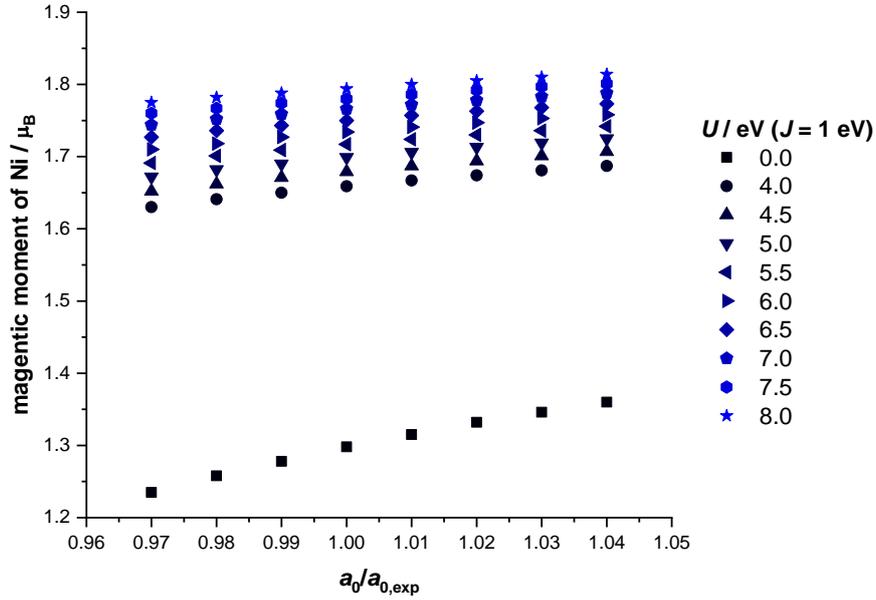

**Figure S1.** The dependence of Ni magnetic moment on the applied value of $U$ and the lattice constant

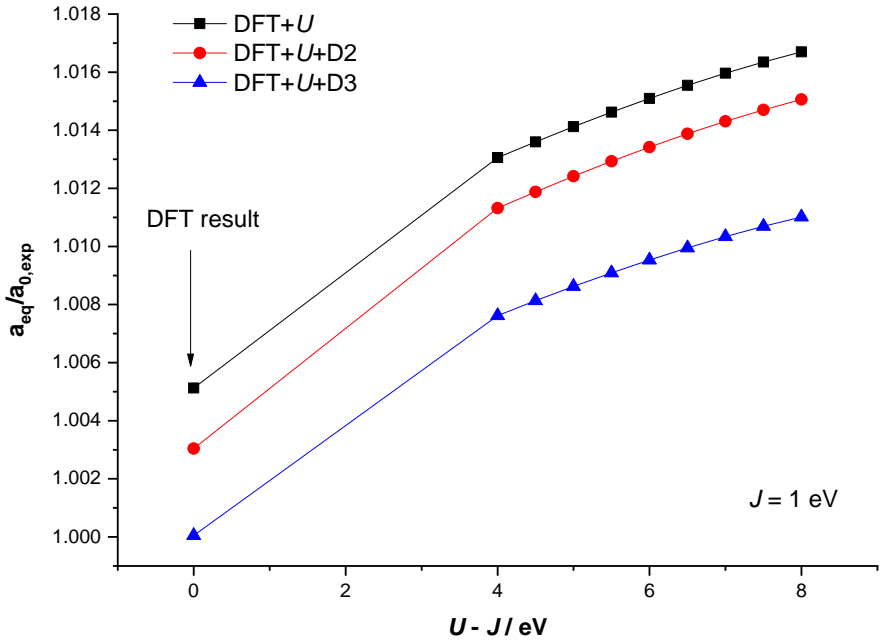

**Figure S2.** The dependence of calculated lattice constant of pristine bulk NiO on the applied $U - J$. For comparison, we also included the data with the dispersion correction (D2 and D3 of Grimme), which both improve lattice constant slightly, causing lattice compression. However, we not that dispersion correction was not further used in this work.

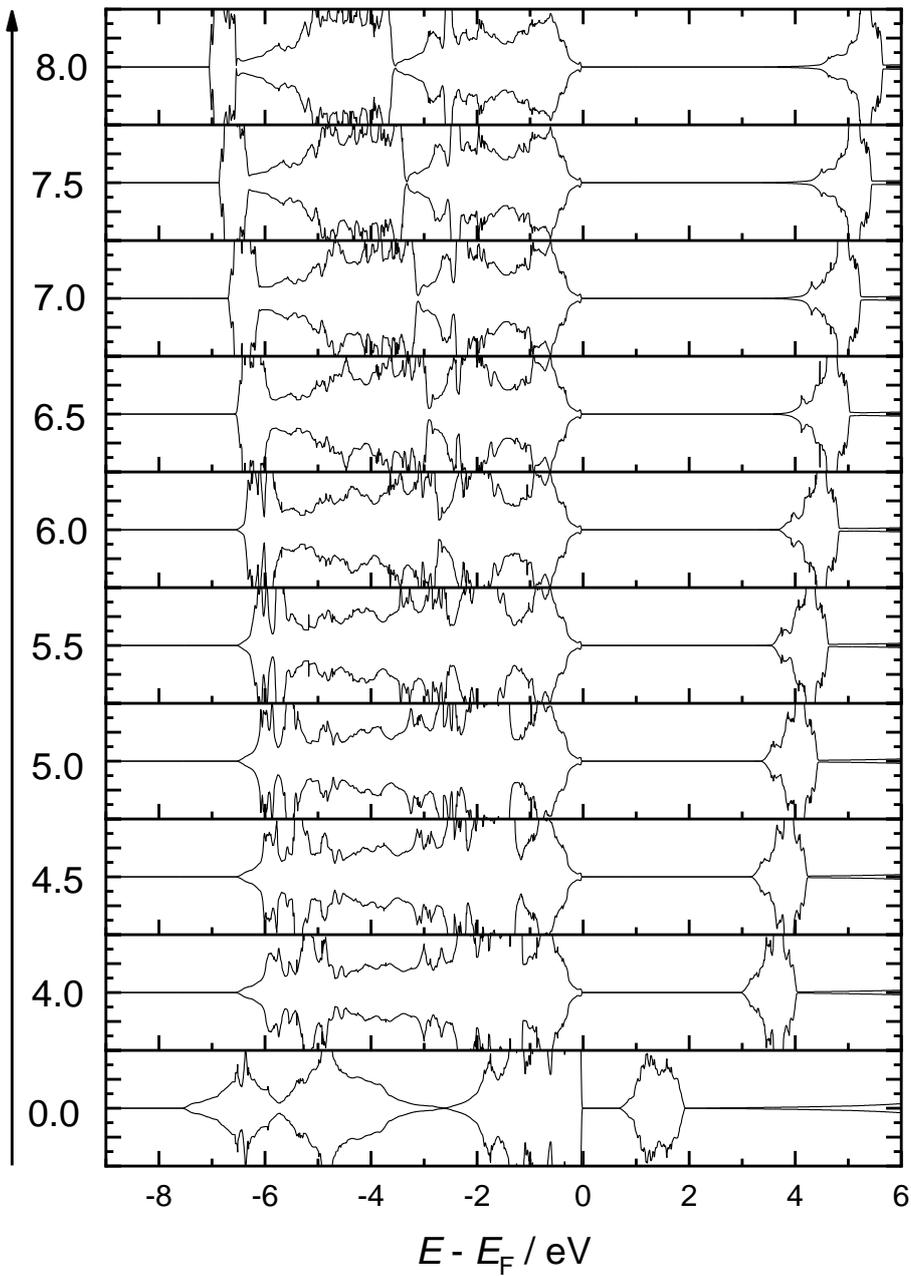

**Figure S3.** Calculated densities of states of pristine bulk NiO for different applied values of $U - J$.

## S2. Optical properties

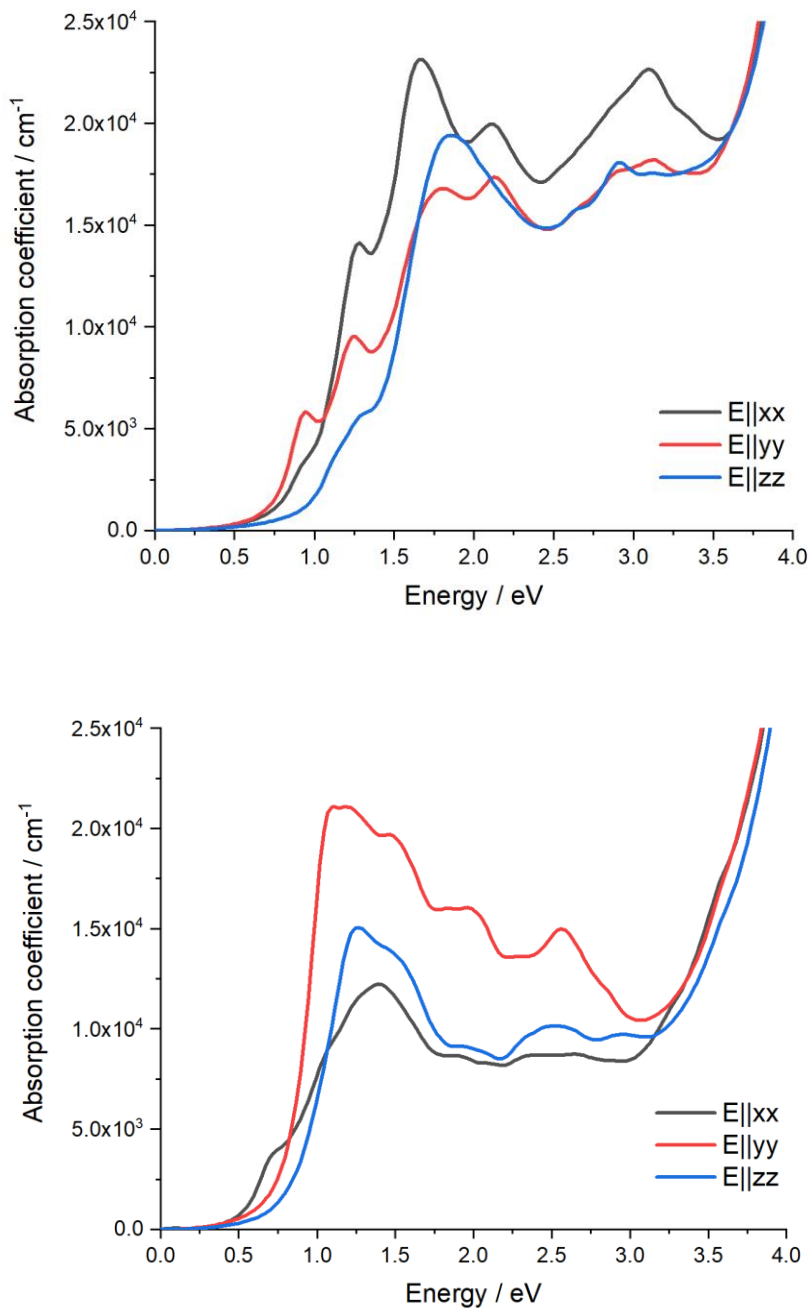

**Figure S4.** Diagonal components of the absorption coefficients of $NiO_{1.019}$ (top) and Li-intercalated in $NiO_{1.019}$ (bottom) with respect to photon energies, as calculated by DFT+$U$.

**S3. Bader analysis**

The following results are obtained for the smaller rhombohedral cell, but identical results are obtained in the larger hexagonal cell. Namely, when the Ni vacancy is formed, Ni atoms bear rather uniform partial charges between +1.28 and +1.29 e. This means that the Ni partial charge is almost the same as in pristine NiO (+1.27 e). Upon insertion of Li into the Ni vacancy, 0.89 e is transferred to the NiO lattice. This charge does not cause a significant change in Ni atoms partial charges as upon Li insertion partial charges of Ni range between +1.26 and +1.27 e. In total, only 0.12 e (out of 0.89) is transferred to Ni atoms.